\documentclass[aps,prb,reprint,superscriptaddress,twocolumn]{revtex4-2}
\usepackage[utf8]{inputenc}
\usepackage{amsmath}
\usepackage{graphicx}
\usepackage{bm}
\usepackage{here}
\usepackage{braket}
\usepackage{physics}
\usepackage{xcolor}
\usepackage{color}
\usepackage[
colorlinks = true,
linkcolor = blue,
citecolor=blue,
urlcolor = blue
]{hyperref}
\graphicspath{{Figure/}}

\begin{document}

\title{Demonstration of highly-sensitive wideband microwave sensing using ensemble nitrogen-vacancy centers}

\author{Kensuke Ogawa}
\email{kensuke.ogawa@phys.s.u-tokyo.ac.jp}
\affiliation{Department of Physics, The University of Tokyo, Bunkyo-ku, Tokyo 113-0033, Japan}
\author{Shunsuke Nishimura}
\affiliation{Department of Physics, The University of Tokyo, Bunkyo-ku, Tokyo 113-0033, Japan}
\author{Kento Sasaki}
\affiliation{Department of Physics, The University of Tokyo, Bunkyo-ku, Tokyo 113-0033, Japan}
\author{Kensuke Kobayasahi}
\email{kensuke@phys.s.u-tokyo.ac.jp}
\affiliation{Department of Physics, The University of Tokyo, Bunkyo-ku, Tokyo 113-0033, Japan}
\affiliation{Institute for Physics of Intelligence, The University of Tokyo, Bunkyo-ku, Tokyo 113-0033, Japan}
\affiliation{Trans-scale Quantum Science Institute, The University of Tokyo, Bunkyo-ku, Tokyo 113-0033, Japan}

\date{\today}

\begin{abstract}
Microwave magnetometry is essential for the advancement of microwave technologies. We demonstrate a broadband microwave sensing protocol using the AC Zeeman effect with ensemble nitrogen-vacancy (NV) centers in diamond. A widefield microscope can visualize the frequency characteristics of the microwave resonator {and the spatial distribution of off-resonant microwave amplitude.} Furthermore, by combining this method with dynamical decoupling, {we achieve the microwave amplitude sensitivity of $5.2 \, \mathrm{\mu T} / \sqrt{\mathrm{Hz}}$, which is 7.7 times better than $40.2 \, \mathrm{\mu T} / \sqrt{\mathrm{Hz}}$ obtained using the protocol in previous research over a sensing volume of $2.77 \, \mathrm{\mu m} \times 2.77 \, \mathrm{\mu m} \times 30 \, \mathrm{nm}$.} Our achievement is a concrete step in adapting ensemble NV centers for wideband and widefield microwave imaging.
\end{abstract}

\maketitle

\addcontentsline{toc}{section}{Introduction}
Over the past decades, microwave has played a significant role in various fields. Microwave engineering \cite{pozar2011microwave}, represented by microwave devices, supports modern communication technology. Coherent spin waves \cite{Pirro2021a} with frequencies in the microwave region are expected to have great potential for application in next-generation information devices. The development of a microwave sensing method is essential for the further development of these microwave-based technologies. In particular, the ability to image local microwave distribution with high sensitivity and broad bandwidth is crucial to device evaluation and visualization of magnetic dynamics.
\par
Nitrogen-vacancy (NV) centers \cite{doherty2013nitrogen} have garnered significant attention as sensors for microwave \cite{Wang2015,wang2022picotesla,Alsid2023}. Taking advantage of the fact that they are sensitive to microwave with frequency around $2.87 \, \mathrm{GHz}$ of their resonance frequency, they have been applied to evaluating microwave devices \cite{Shao2016,Horsley2018} and quantitative visualization of coherent spin waves in magnetic materials \cite{van2015nanometre,Andrich2017,kikuchi2017long,bertelli2020magnetic,zhou2021magnon,Carmiggelt2023}. 
\par
In most studies, microwave sensing using NV centers is based on measuring Rabi oscillation. This approach, however, is inherently limited in bandwidth; the sensitivity degrades rapidly as the frequency detunes from the resonance frequency. While the detectable frequency range can be tuned by adjusting the resonance frequency through changing the static magnetic field strength, this technique can be invasive for some targets, especially for materials with magnetic properties that are dependent on the magnetic field. \par
To expand the detection bandwidth without changing the static magnetic field, a protocol for detecting the AC Zeeman effect \cite{autler1955stark,ramsey1955resonance,wei1997strongly} has been proposed \cite{li2019wideband}. The AC Zeeman effect refers to a slight shift in the resonance frequency of the NV center under microwave irradiation. By detecting this shift, it has been demonstrated that off-resonant microwave with a detuning of approximately $1 \, \mathrm{GHz}$ can be measured using a single NV center \cite{li2019wideband}. This effect has also been observed in superconducting qubits \cite{schneider2018local} and adapted for evaluating the transmission characteristics of microwave circuits \cite{kristen2020amplitude}. \par
Although the utility of the protocol based on detecting the AC Zeeman effect has been demonstrated in proof-of-principle experiments, its application to the characterization of practical systems including microwave circuits remains unexplored. Additionally, to expand its use in widefield imaging requires a protocol that maintain high sensitivity, even when utilizing ensemble NV centers which typically have shorter coherence time, $T_{2}$. \par
In this paper, {we use a widefield microscopy} and ensemble NV centers to detect the AC Zeeman effect. We show that it can be applied to practical systems by obtaining the frequency response of a microwave planar ring antenna. {Also, we demonstrate the visualization of the spatial distribution of off-resonant microwave amplitude on an omega-shaped antenna.} In addition, we develop sequences that utilize dynamic decoupling to improve the coherence of the NV ensemble and achieve a sensitivity improvement of 7.7 times compared to a conventional sequence. The results of our study serve as a significant step in applying NV centers to widefield and wideband microwave sensing. \par

\addcontentsline{toc}{section}{Experimental Setup and Principles}
First, we describe the principle of microwave sensing based on the AC Zeeman effect. The AC Zeeman effect refers to a phenomenon in which the resonance frequency is shifted by the irradiation of microwave \cite{autler1955stark,ramsey1955resonance,wei1997strongly}. {The shift $f_{\text{ACZ}}$ can be expressed as $f_{\text{ACZ}} = \sqrt{\Delta^2 + \Omega^2} - \Delta$}, where $\Omega = \frac{\gamma_{e} B_\text{mw}}{\sqrt{2}}$ is the Rabi frequency of NV center, {which is proportional to} the microwave amplitude $B_\text{mw}$, $\gamma_{e}$ is the gyromagnetic ratio of an electron spin, and {$\Delta = f_{\mathrm{NV}} - f_{\mathrm{mw}}$ is the detuning of microwave frequency $f_{\mathrm{mw}}$ from the resonance frequency of NV center $f_{\mathrm{NV}}$.} Under the condition where $\Delta \gg \Omega$, {the shift $f_{\mathrm{ACZ}}$ can be approximated as} \par

\begin{equation}
    f_{\mathrm{ACZ}} \approx \frac{\Omega^2}{2\Delta} = \frac{(\gamma_e B_\text{mw})^2}{4\Delta}.
    \label{eqfacz}
\end{equation}

\begin{figure}
    \centering
    \includegraphics[width=\linewidth]{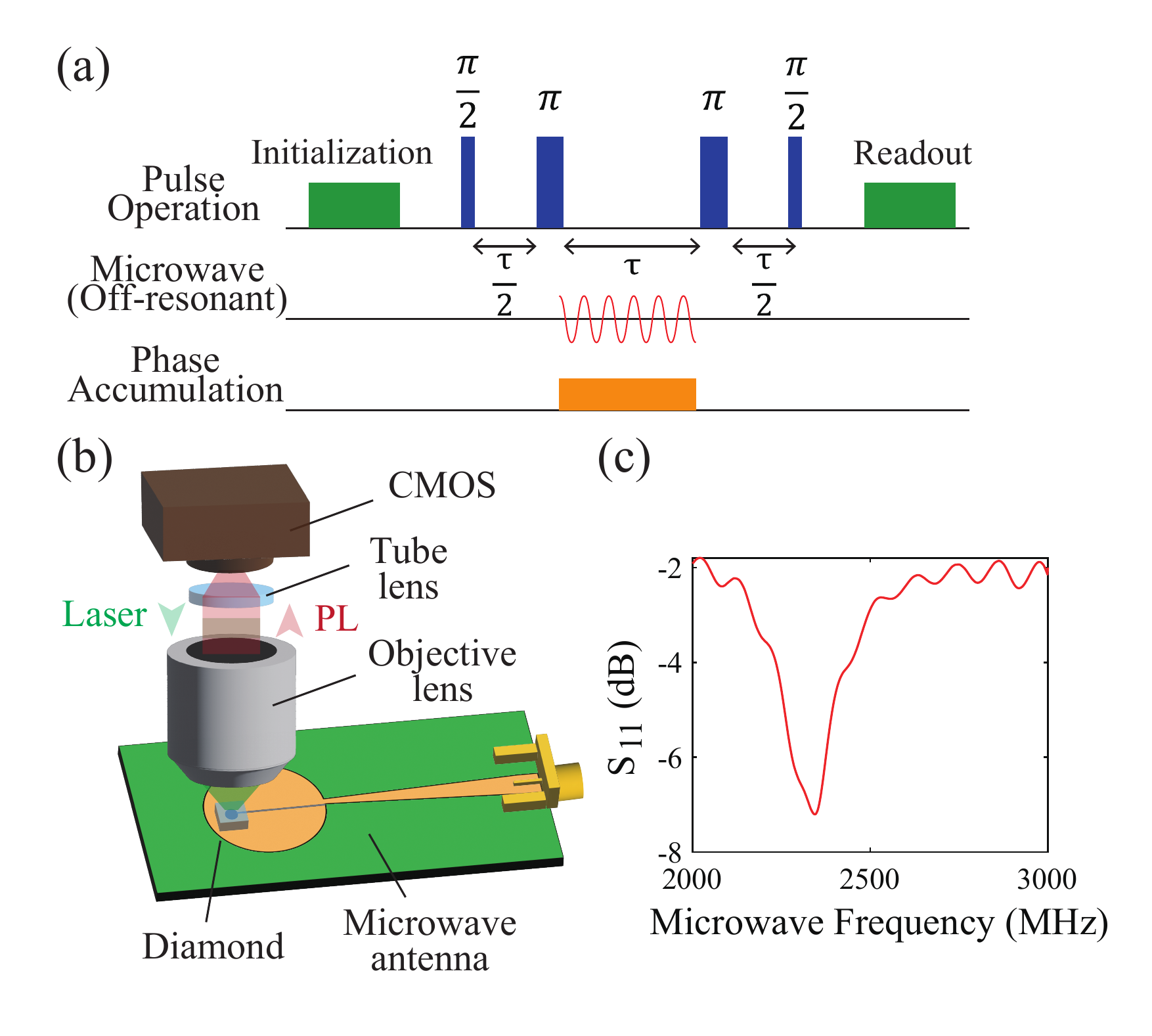}
    \caption{(a) Pulse sequence for detecting the AC Zeeman effect. The green color pulse corresponds to laser and the purple color pulse to resonant microwave. (b) Schematic illustration of the experimental setup. The diamond containing ensemble NV centers is placed on the planar ring antenna \cite{Sasaki2016,Misonou2020}. (c) $S_{11}$ of the antenna.}
    \label{fig:1}
\end{figure}

 This shift can be precisely quantified through the pulse sequence shown in Fig. \ref{fig:1}(a), which was developed in previous work \cite{li2019wideband}. After initialization by the green laser, the Carr-Purcell 2 ($\mathrm{CP2}$) sequence \cite{carr1954effects} is applied by resonant microwave followed by the reading out of the spin state. Off-resonant microwave is applied only for the time between the two $\pi$ pulses to accumulate phase due to the AC Zeeman effect. Although the Ramsey fringe measurement can also detect the AC Zeeman effect, such spin echo type measurement is effective in suppressing low-frequency magnetic field noise. The obtained photoluminescence (PL) intensity normalized by that of the initial state can be written as

\begin{equation}
     S(\tau) = 1 - \frac{1 - \cos \qty(2\pi f_{\text{ACZ}} \tau) e^{\frac{-2\tau}{T_{2}}}}{2} C, 
     \label{eqsig}
\end{equation}

where $T_{2}$ is the coherence time under the $\mathrm{CP2}$ sequence, and $C$ is a PL contrast between the $m_{\text{s}} = 0$ and $m_{\text{s}} = -1$ state. A rigorous expression for the time evolution is given in supplementary material. \par
The experimental setup is shown in Fig. 1(b). We use a homemade widefield fluorescence microscope \cite{ogawa2023lock}. A green laser of $515 \, \mathrm{nm}$ wavelength is irradiated, and the red emission from NV centers is detected by a CMOS camera. We use ensemble NV centers near the surface (depth\,$\sim30$~nm) of the electronic grade I\hspace{-1.2pt}Ia diamond (Element Six) created by $\mathrm{^{15}N}^+$ ion implantation (10~keV, $1\times10^{13}~\text{cm}^{-2}$) and subsequent annealing \cite{Tetienne2018,Sangtawesin2019,healey2020comparison}. To resolve the degeneracy in the NV center's spin energy level, a static magnetic field of about $11 \, \mathrm{mT}$ is applied in the (111) direction using a permanent magnet. We use $m_{\text{s}} = 0 \ \text{and} -1$ states as an effective two-level system {whose resonance frequency $f_{\mathrm{NV}}$ is around $2560 \, \mathrm{MHz}$.} {A planar ring antenna \cite{Sasaki2016,Misonou2020} is used as the measurement target except for an imaging measurement.} It functions as an LC resonant circuit; the central hole serves as an inductor, and the gap as a capacitor. The reflection characteristics of the antenna $S_{11}$ measured with a vector network analyzer (VNA) is shown in Fig. \ref{fig:1}(c). It has a resonance frequency around $2370 \, \mathrm{MHz}$. Since it has a wide bandwidth, we also use this antenna for irradiation of resonant microwave pulses. Generated microwaves are concentrated near the center of the hole, resulting in uniform microwave irradiation in the field of view. \par

\begin{figure}
    \centering
    \includegraphics[width=\linewidth]{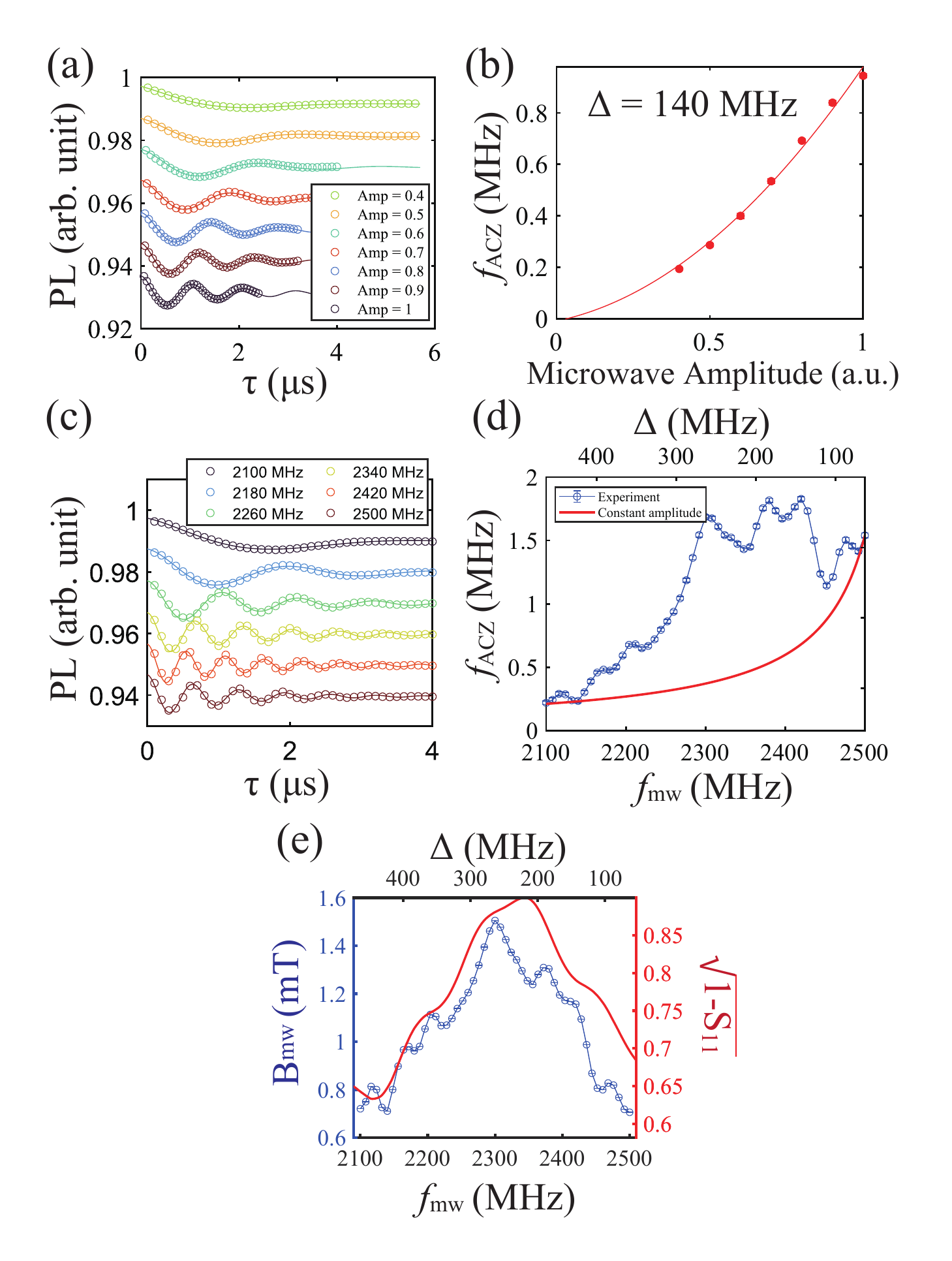}
    \caption{Measurement of microwave antenna characteristics using the AC Zeeman effect. (a) PL contrasts on off-resonant microwave sweep time $\tau$ at several microwave amplitudes. Off-resonant microwave frequency {$f_{\text{mw}}$} is fixed at $2420 \, \mathrm{MHz}$. (b) Microwave amplitude dependence of signal frequency {$f_{\text{ACZ}}$}. A quadratic function is used for fitting [see Eq. (\ref{eqsig})]. (c) Signal dependence on microwave sweep time for different off-resonant microwave frequencies. (d) {Signal frequency $f_{\text{ACZ}}$ dependence on off-resonant microwave frequency $f_{\text{mw}}$. Blue points are experimental results, and red line is expected signal frequencies $f_{\text{ACZ}}$  when the microwave amplitude is constant at the value of $f_{\text{mw}} = 2500 \, \text{MHz}$.} (e) {Dependence of microwave amplitude {$B_{\mathrm{mw}}$} on the microwave frequency {$f_{\text{mw}}$}. Blue points are calculated from the signal frequencies {$f_{\text{ACZ}}$} and the detuning {$\Delta$} using Eq. (\ref{eqfacz}). Red line is estimated microwave amplitude calculated from the reflection characteristics [Fig. \ref{fig:1}(c)].}}
    \label{fig:2}
\end{figure}

\addcontentsline{toc}{section}{Proof-of-principle experiment and microwave circuit evaluation}
We present experimental results demonstrating the detection of the AC Zeeman effect on ensemble NV centers as a proof-of-principle expetiment. Figure~\ref{fig:2}(a) is the result of the detection of the off-resonant microwave with a frequency of $2420 \, \mathrm{MHz}$, detuned by approximately $140 \, \mathrm{MHz}$ away from the resonance frequency. Each curve is the result of varying the amplitude of the off-resonant microwave input to the antenna. The PL contrast oscillates due to phase accumulation caused by the AC Zeeman effect and the oscillation frequency increases as the microwave amplitude increases. The signal decay time is approximately $1.6 \, \mathrm{\mu s}$, which corresponds to half the time of $T_{2}$ under the $\mathrm{CP2}$ sequence. The solid lines are the results of the fitting using Eq. (\ref{eqsig}). The signals are consistent with theoretical predictions at all microwave amplitudes. Figure~\ref{fig:2}(b) plots the signal frequency {$f_{\text{ACZ}}$} for each microwave amplitude. As predicted by Eq.~(\ref{eqsig}), the signal frequency follows the square of the amplitude [solid line in Fig.~\ref{fig:2}(b)]. \par
Next, we demonstrate the quantification of the resonance characteristics of the microwave antenna. {We sweep the off-resonant microwave frequency $f_{\mathrm{mw}}$ from $2200 \, \mathrm{MHz}$ to $2500 \, \mathrm{MHz}$ which corresponds to from 360 MHz to 60 MHz in regard to the detuning $\Delta$} while fixing its amplitude. The obtained PL contrasts are shown in Fig. \ref{fig:2}(c). {The oscillation frequency decreases as the detuning increases.} The resulting signal frequency {$f_{\text{ACZ}}$} at each microwave frequency {$f_{\text{mw}}$} is shown in Fig. \ref{fig:2}(d). {The blue points correspond to the experimental result, and the red line shows the expected signal frequency $f_{\text{ACZ}}$ when the microwave amplitude is independent of microwave frequency $f_{\text{mw}}$ and constant at the value of $f_{\text{mw}} = 2500 \, \text{MHz}$ for reference.} Since the antenna has its own resonance characteristics [see Fig.~\ref{fig:1}(c)], {the blue points do not follow the red line}. Using Eq.~(\ref{eqfacz}), we estimate the amplitude $B_{\mathrm{mw}}$ at each microwave frequency. The result is shown in Fig. \ref{fig:2}(e). The blue {points} represents the converted microwave amplitude $B_{\mathrm{mw}}$. The red line represents the estimated microwave amplitude calculated from the reflection characteristics gained by VNA for comparison [see Fig. \ref{fig:1}(c)]. The good similarity of these results confirms that the frequency dependence of the antenna can be accurately measured. Notably, compared to the characteristics seen in the VNA measurement (red line), the extracted microwave amplitude is subject to periodic modulation. This can be due to the formation of standing waves in the whole microwave circuit, including the cable, circulator, and amplifier. Our result indicates that it is realistically possible to directly visualize the frequency dependence of microwave applied to a sample.

\addcontentsline{toc}{section}{Imaging off-resonant microwave using CP2 sequence}

\begin{figure}
    \centering
    \includegraphics[width=\linewidth]{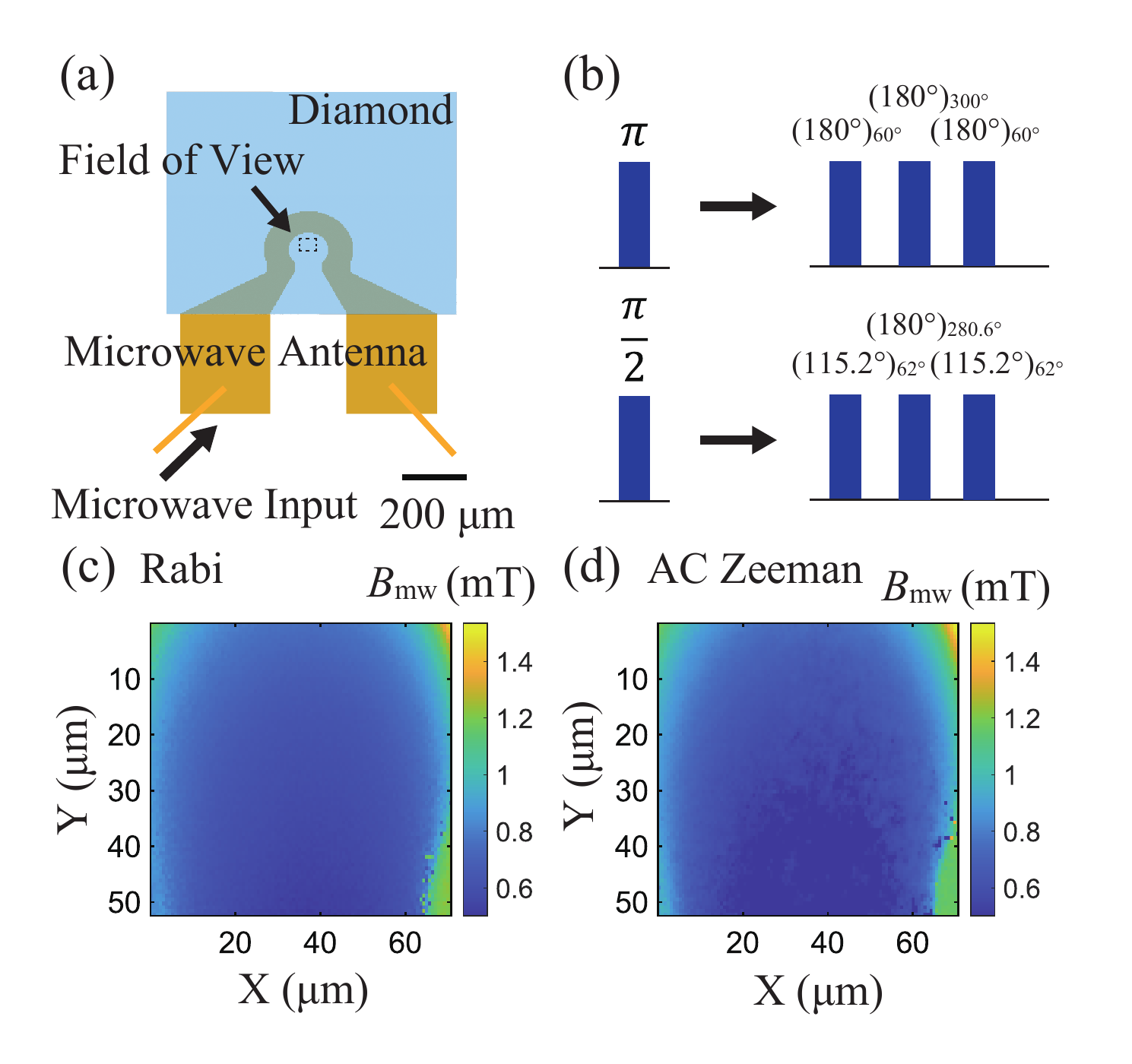}
    \caption{{Imaging off-resonant microwave distribution using the AC Zeeman effect. (a) Schematic illustration
of the experimental setup for the imaging measurement. The diamond sample containing ensemble NV centers is placed on the omega-shaped antenna. (b) Illustrations of SCROFULOUS composite pulse sequences corresponding to $\pi$ and $\frac{\pi}{2}$ pulses. The angle and the subscript angle correspond to the rotation angle and the phase, respectively. (c) Distribution of the resonant microwave amplitude obtained by a Rabi oscillation measurement. (d) Distribution of the off-resonant microwave amplitude with the detuning $\Delta = 150 \, \mathrm{MHz}$ obtained using the AC Zeeman effect.}}
    \label{fig:4}
\end{figure}

We then demonstrate the imaging measurement. The experimental setup is depicted in Fig. \ref{fig:4}(a). For the imaging measurement, instead of using the planar ring antenna shown in Fig. \ref{fig:1} (b), to create a distribution of microwave amplitude within the field of view, we employ an omega-shaped microwave antenna with an outer diameter of $250 \, \mathrm{\mu m}$ and an inner diameter of $100 \, \mathrm{\mu m}$ made of copper with a thickness of $300 \, \mathrm{nm}$ on a silicon substrate. Both resonant microwave pulses and off-resonant signal microwave are output from the antenna. For the control pulses, to reduce pulse length errors caused by the distribution of microwave amplitude within the field of view, we use composite pulses, particularly SCROFULOUS composite pulse sequence \cite{cummins2003tackling,nomura2021composite} (see supplementary material for details). SCROFULOUS sequence suppresses pulse length errors by combining three pulses with distinct nominal rotation angles and phases. In Fig. \ref{fig:4}(b), we present SCROFULOUS sequences corresponding to $\pi$ and $\frac{\pi}{2}$ pulses.\par
\begin{figure}
    \centering
    \includegraphics[width=\linewidth]{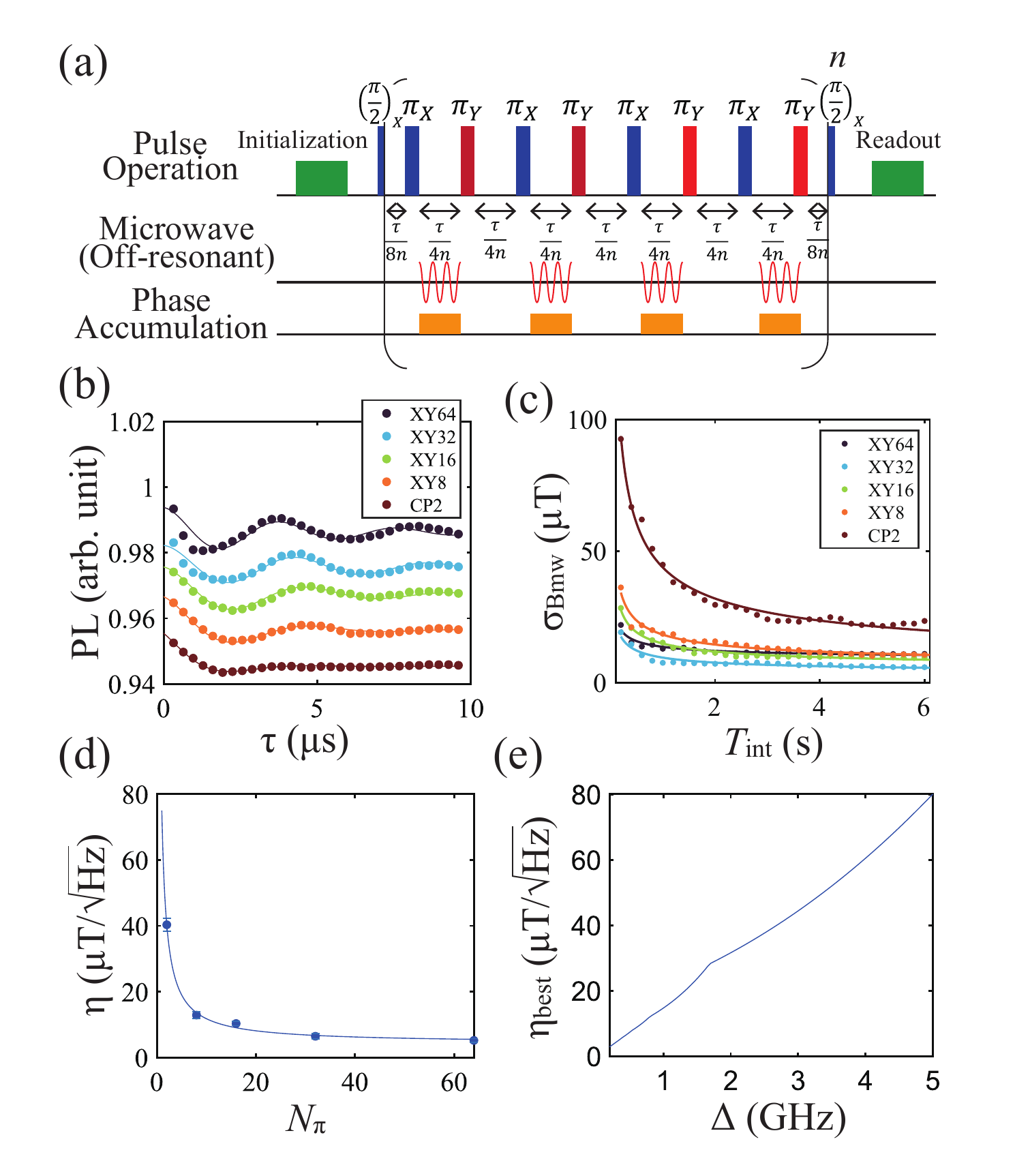}
    \caption{Measurement of the AC Zeeman effect using more improved pulse sequences. (a) Pulse sequence. For resonant microwave, the $\mathrm{XY8}$ sequences are repeated $n$ times. For off-resonant signal microwave, it is applied once every two times between $\pi$ pulses. (b) Obtained PL contrasts on different pulse sequences. (c) Integration time {$T_{\text{int}}$} scaling of {the standard error $\sigma_{B_{\mathrm{mw}}}$} of microwave amplitude $B_{\mathrm{mw}}$. (d) Relationship between the number of $\pi$ pulses {$N_{\pi}$} and the sensitivity $\eta$ of $140 \, \mathrm{MHz}$ detuned microwave. (e) Simulation of the detuning $\Delta$ dependence of best sensitivity $\eta_{\text{best}}$ at a microwave amplitude of $B_{\mathrm{mw}} = 0.75 \, \mathrm{mT}$.}
    \label{fig:3}
\end{figure}

First, the spatial distribution of resonant microwave amplitude obtained by Rabi oscillation measurement is displayed in Fig. \ref{fig:4}(c). A clear distribution of microwave amplitude within the field of view is observed, with a difference of about threefold between the maximum and minimum values. Next, in Fig. \ref{fig:4}(d), we show the result of imaging the amplitude of off-resonant microwave with detuning $\Delta = 150 \, \mathrm{MHz}$ by combining the protocol shown in Fig. \ref{fig:1}(a) with SCROFULOUS composite pulse sequences [Fig. \ref{fig:3}(b)]. The observed microwave amplitude distribution is consistent with the distribution measured with Rabi oscillation. This conformity confirms the feasibility of imaging off-resonant microwave amplitude using the AC Zeeman effect.

\addcontentsline{toc}{section}{Sensing with more advanced pulse protocols}
Until now, we have demonstrated a visualization of the characteristics of a microwave antenna by detecting the AC Zeeman effect on ensemble NV centers. As the usefulness of this protocol has been confirmed, next we examine its applicability: as Eq. (\ref{eqsig}) shows, the signal frequency is proportional to the detuning inversely and to the square of the amplitude, hence microwave with large detuning and small amplitude would make the signal frequency very small. When attempting imaging measurements with ensemble NV centers that have a short coherence time, sufficient sensitivity may not be obtained because the signal decays before it oscillates for a full cycle. To mitigate this problem, we introduce an improved protocol by combining an $\mathrm{XY8}$ sequence, which can extend coherence time. \par
The pulse protocol is shown in Fig. \ref{fig:3}(a). $\mathrm{XY8}$ sequences \cite{gullion1990new} are repeated $n$ times ($n = 1, 2, 4, \text{and} \ 8$). Applying $\pi$ pulses more frequently increases the filter frequency that cancels the effects of magnetic noise and thus extends coherence time \cite{joas2017quantum}. We set the duration of the XY sequence to be $2 \tau$, same as the CP2 sequence shown in Fig.~\ref{fig:1}(a), and redefined the pulse interval as $\frac{\tau}{4n}$. We apply off-resonant microwave once every two $\pi$ pulses to accumulate phase induced by the AC Zeeman effect. \par
We show how much our sequence [Fig.~\ref{fig:3}(a)] improves sensitivity. The obtained PL contrast for each sequence are shown in Fig. \ref{fig:3}(b). The NV centers' resonance frequency {$f_{\mathrm{NV}}$} is $2560 \, \mathrm{MHz}$, and the off-resonant microwave frequency {$f_{\mathrm{mw}}$} is fixed at $2420 \, \mathrm{MHz}$. As expected, as the number of $\pi$ pulses $N_{\pi} (= 8n)$ increases, the signal decay becomes slower due to the extension of $T_{2}$. For sequences with many pulses, such as $\mathrm{XY32}$ or $\mathrm{XY64}$, the signals are slightly distorted in the region where $\tau$ is small, say $\tau < 1 \, \mathrm{\mu}$s. We numerically confirm that the signal can be a little modulated by applying a large number of $\pi$ pulses (see supplementary material for details). Also, this distortion can be due to the limited bandwidth of microwave antenna because off-resonant microwave pulses are finely segmented in these sequences. \par
We examine the protocol dependence of the sensitivity of microwave field amplitude $B_{\mathrm{mw}}$. To evaluate sensitivity, we use PL contrasts at $10 \times 10$ imaging pixels corresponding to $(2.77 \, \mathrm{\mu m})^2$ within the field of view. Figure \ref{fig:3}(c) shows the signal integration time $T_{\text{int}}$ dependence of {the standard error} $\sigma_{\mathrm{B_{\mathrm{mw}}}}$ of the off-resonant microwave amplitude. Details of the estimation method of $\sigma_{B_\text{mw}}$ are described in supplementary material. {The standard error} decreases with integration time for all pulse sequences and the decrease is significantly faster when the pulse number is large. {The standard error} can be separated into the statistical error due to the signal’s shot noise, which is inversely proportional to the square root of the integration time and the systematic error $\sigma_{0}$, as expressed in the following equation,
\begin{equation}
    \sigma_{B_{\mathrm{mw}}}(T) = \eta T^{-0.5} + \sigma_{0},
    \label{eq:sens_scale}
\end{equation}
where $\eta$ corresponds to the sensitivity of microwave field amplitude. Figure \ref{fig:3}(d) shows the results of the sensitivity $\eta$ obtained by fitting the results in Fig. \ref{fig:3}(c) using Eq. (\ref{eq:sens_scale}). It quantitatively confirms that the sensitivity improves as the number of pulses increases. In particular, the sensitivity of the $\mathrm{XY64}$ sequence is {$5.22 \, \mathrm{\mu T} / \sqrt{\mathrm{Hz}}$}, which is 7.7 times better compared to the sensitivity of the $\mathrm{CP2}$ sequence of {$40.2 \, \mathrm{\mu T} / \sqrt{\mathrm{Hz}}$.} \par
Next, we discuss the sensitivity scaling with respect to the number of pulses. The scaling behavior is evaluated by fitting the result in Fig. \ref{fig:3}(d) with
\begin{equation}
  \eta = \eta_{0} N_{\pi}^{-p}.
\end{equation}

We obtain the scaling factor as $p = 0.98$. This value depends on the scaling of $T_{2}$ and the sampling conditions of the microwave sweep time (see supplementary material for details). \par
Finally, we discuss the applicability of the improved protocol based on the best sensitivity $\eta_{\mathrm{best}}$. The best sensitivity can be obtained by measuring at only one fixed microwave sweep time $\tau$.
The sensitivity $\eta_{\mathrm{single}}$ when measuring at only one fixed $\tau$ can be written as
\begin{equation}
    \eta_{\mathrm{single}} = \frac{\sqrt{2} \Delta \sigma_{\mathrm{s}} \sqrt{2\tau + \tau_{\mathrm{read}}}}{\gamma_{e} \pi \Omega \tau C \left| \sin \qty(\pi \frac{\Omega^2}{\Delta}\tau) \right| e^{-\frac{-2\tau}{T_{2}}}},
    \label{eq:etas}
\end{equation}
where $\tau_{\mathrm{read}} = 64 \, \mathrm{\mu s}$ is the readout time due to laser irradiation and $\sigma_{\mathrm{s}}$ is the signal noise in one readout (see supplementary material for details). The best sensitivity $\eta_{\mathrm{best}}$ in this measurement is given by
\begin{equation}
    \eta_{\mathrm{best}} = \displaystyle \min_{\tau} \qty(\eta_{\mathrm{single}}).
\end{equation}
Figure \ref{fig:3}(e) shows the simulation results of sensitivity $\eta_{\mathrm{best}}$ of $\mathrm{XY64}$ protocol at different detunings under the microwave amplitude $B_{x} = 0.75 \, \mathrm{mT}$, which is typical microwave amplitude from spin waves \cite{bertelli2020magnetic, zhou2021magnon}, and other parameters in our experiment. Although the sensitivity degrades as the detuning increases, even at a detuning of $5 \, \mathrm{GHz}$, it has a sensitivity of approximately $84 \, \mathrm{\mu T / \sqrt{Hz}}$. This means that measurement with a resolution of $1.7 \, \mathrm{MHz}$ in terms of Rabi frequency can be attained in one second. The sensitivity of our sequence can be further enhanced by at least an order of magnitude by optimizing the experimental setup (see supplementary material for details).

\addcontentsline{toc}{section}{Conclusion}
In conclusion, we have demonstrated off-resonant microwave sensing by detecting the AC Zeeman effect on ensemble NV centers using a widefield microscope. We visualized the frequency characteristics of a planar ring antenna, {and the spatial distribution of off-resonant microwave on an omega-shaped antenna.} We also developed a protocol in combination with $\mathrm{XY8}$ sequences, demonstrating that the sensitivity can be enhanced up to 7.7 times by extending coherence time. Our technique holds promise for imaging microwave in tens of GHz range on a micrometer scale without modulating the magnetic field by optimizing diamond samples and measurement conditions. Such capabilities pave the way for directly imaging coherent spin waves, such as those in van der Waals magnets, which are far detuned from the resonance frequency of NV centers at low magnetic field \cite{shen2021multi}. Overall, this work serves as a foundational step in adapting the ensemble NV centers for practical wideband and widefield microwave imaging. \par

\vskip\baselineskip

\addcontentsline{toc}{section}{Supplementary Material}
See supplementary material for details of experimental setup, the AC Zeeman effect and time evolution by pulse sequence, composite pulse sequences, sensitivity estimation, discussions about scaling factor of the sensitivity's dependence on the number of $\pi$ pulses, strategies for optimizing the experimental setup to enhance sensitivity, and comparison of the sensitivity with previous research.

\vskip\baselineskip

\addcontentsline{toc}{section}{Acknowledgments}
This work was partially supported by ``Advanced Research Infrastructure for Materials and Nanotechnology in Japan (ARIM)" (Proposal No. JPMXP1222UT1131) of the Ministry of Education, Culture, Sports, Science and Technology of Japan (MEXT), {Kondo Memorial Foundation}, Grants-in-Aid for Scientific Research (Nos. JP22K03524, JP23H01103, JP19H00656, and JP19H05826), and {JST, CREST Grant Number JPMJCR23I2, Japan.} {S.N and K.O. acknowledges financial support from FoPM, WINGS Program, the University of Tokyo.} K.O. is supported by Grant-in-Aid for JSPS Fellows (Nos. JP22KJ1058). {S.N. acknowledges support from JSR Fellowship, the University of Tokyo.}

\end{document}


\title{Supplementary Material: Demonstration of highly-sensitive wideband microwave sensing using ensemble nitrogen-vacancy centers}

\author{Kensuke Ogawa}
\email{kensuke.ogawa@phys.s.u-tokyo.ac.jp}
\affiliation{Department of Physics, The University of Tokyo, Bunkyo-ku, Tokyo 113-0033, Japan}
\author{Shunsuke Nishimura}
\affiliation{Department of Physics, The University of Tokyo, Bunkyo-ku, Tokyo 113-0033, Japan}
\author{Kento Sasaki}
\affiliation{Department of Physics, The University of Tokyo, Bunkyo-ku, Tokyo 113-0033, Japan}
\author{Kensuke Kobayasahi}
\email{kensuke@phys.s.u-tokyo.ac.jp}
\affiliation{Department of Physics, The University of Tokyo, Bunkyo-ku, Tokyo 113-0033, Japan}
\affiliation{Institute for Physics of Intelligence, The University of Tokyo, Bunkyo-ku, Tokyo 113-0033, Japan}
\affiliation{Trans-scale Quantum Science Institute, The University of Tokyo, Bunkyo-ku, Tokyo 113-0033, Japan}

\date{\today}
\maketitle

\renewcommand{\thefigure}{S\arabic{figure}}
\renewcommand{\theequation}{S\arabic{equation}}

\section*{A. Experimental Details}
\subsection*{Apparatus for pulse measurement and image acquisition protocol}
\begin{figure}
    \centering
    \includegraphics[width=\linewidth]{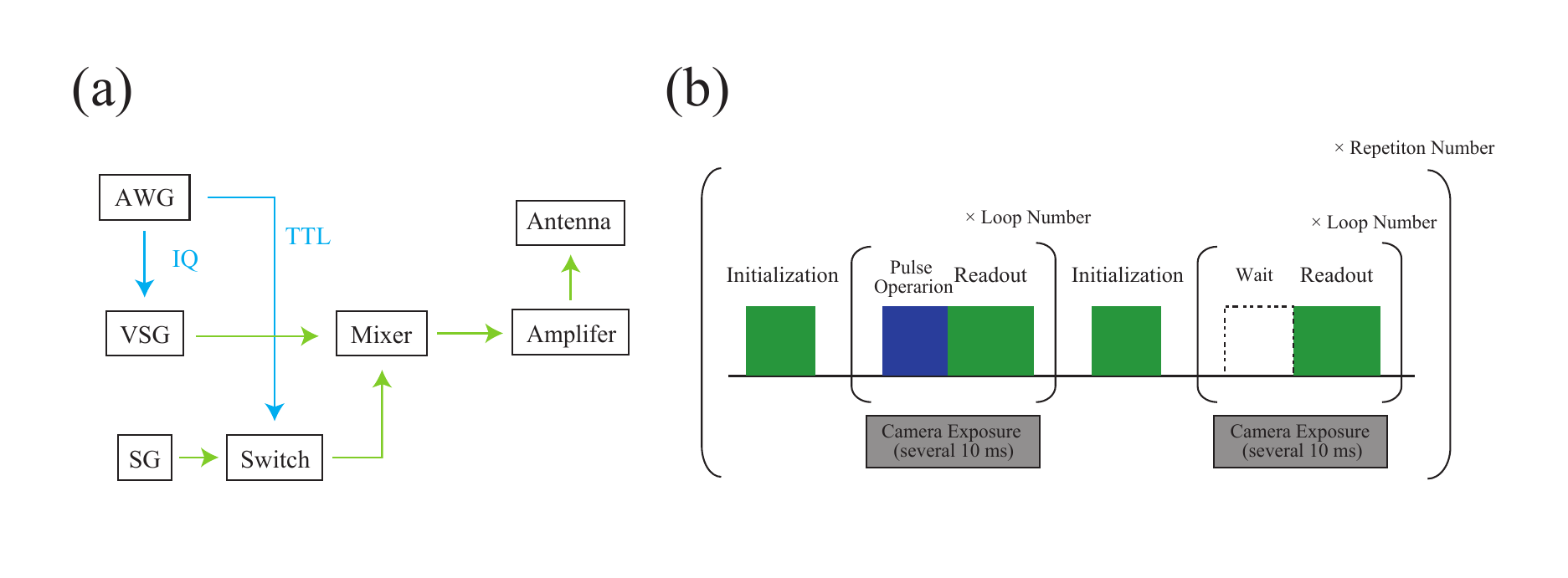}
    \caption{(a) Details of microwave application method. (b) Protocol for optical detection using CMOS camera.}
    \label{fig:sa}
\end{figure}

We input off-resonant signal microwave and resonant microwave for pulse operation into the antenna. The setup is shown in Fig. \ref{fig:sa}(a). Microwave is output from two signal generators, a Vector Signal Generator (VSG, ROHDE \& SCHWARZ SMU200A) and a Signal Generator (SG, KEYSIGHT E8257D), both controlled by an Arbitrary Wave Generator (AWG, SPECTRUM M4i.6631-x8). The amplitude and phase of the microwave generated by VSG are controlled by IQ modulation, and the microwave by SG is constantly output and switched on and off by inputting a TTL signal from the AWG to a microwave switch (Mini-Circuits ZYSWA-2-50DR+). The outputs from the two signal generators are mixed by a microwave mixer (Mini-Circuits ZFRSC-42-S+) and amplified by 45 dB before input to the microwave antenna. In the off-resonant microwave amplitude dependence experiment, off-resonant microwave is output from VSG, and resonant microwave for pulse manipulation is output from SG. For subsequent frequency sweep measurements and $\mathrm{XY8}$ sequences measurements, resonant microwave is output from VSG, and off-resonant microwave is output from SG. \par
Figure \ref{fig:sa}(b) shows the signal acquisition protocol used in this study with a CMOS camera (Basler ace acA2040-55um). The basic protocol is based on previous studies \cite{Horsley2018,mariani2020system}. After initializing the NV center with a laser pulse ($64 \, \mathrm{\mu s}$), the camera acquisition begins, followed by microwave pulse operation and readout and initialization with a laser pulse ($64 \, \mathrm{\mu s}$). This sequence is repeated until the camera exposure time. In addition, a signal without pulse sequence is also acquired as a reference signal. We treat the ratio of these two signals as photoluminescence (PL) contrast. \par
The microscope magnification is 50x, and the field of view is $106 \, \mathrm{\mu m} \times 140 \, \mathrm{\mu m}$ (512 px $\times$ 384 px). All signals from this field of view are integrated together except for the sensitivity evaluation. For the sensitivity evaluation, we analyze only the data from $10 \times 10$ pixels.

\section*{B. AC Zeeman effect and time evolution by pulse sequence}
We derive the AC Zeeman shift and time evolution by pulse sequence when off-resonant microwave is applied to an NV center. For simplicity, we consider a two-level system. an NV center can be regarded as a two-level system of $m_{\text{s}} = 0$ and $m_{\text{s}} = -1$ states if a magnetic field to some extent is applied parallel to in the axial direction ({z-axis}).

\subsection*{Derivation of AC Zeeman effect}
The system Hamiltonian $\hat{H}$ when microwave is applied can be written as
\begin{equation}
  \hat{H} = \gamma_{e} B_{0} \hat{S_{z}} + \gamma_{e} B_{\mathrm{mw}} \cos \qty(2 \pi f_{\mathrm{mw}} t + \phi) \hat{S_{x}}. \\
  \label{eqa1}
\end{equation}
Let the eigenstates of $\hat{S_{z}}$ as $\Ket{+}$, $\Ket{-}$ and Eq. (\ref{eqa1}) can be described as
\begin{equation}
    \hat{H} = \frac{\gamma_{e} h}{2} \mqty(B_{0} & B_{\mathrm{mw}} \cos(2\pi f_{\text{mw}} t + \phi) \\
              B_{\mathrm{mw}} \cos(2\pi f_{\text{mw}} t + \phi) & -B_{0}).
\end{equation}
The state $\psi$ of the system evolves in time according to the schr\"odinger equation
\begin{equation}
    i h \frac{d\psi}{dt} = 2\pi \hat{H} \psi.
\end{equation}
When we apply a unitary transformation $\hat{U}$ to $\psi$ and transfer it to the rotational coordinate with $\tilde{\psi} = \hat{U} \psi$, $\tilde{\psi}$ evolves according to the following effective Hamiltonian $\hat{H'}$ can be written as 
\begin{equation}
    \hat{H'} = \hat{U} \hat{H} {\hat{U}}^{\dag} - i \hat{U} {\dot{U}}^{\dag},
\end{equation}
where $\hat{U}$ is defined as
\begin{equation}
    \hat{U} = \exp(i 2\pi f_{\mathrm{mw}} \hat{S_{z}}t). 
\end{equation}
Using the rotational wave approximation, $\hat{H'}$ can be written as
\begin{equation}
    \hat{H'} = \frac{h}{2} \mqty(\Delta & \Omega e^{-i\phi} \\ \Omega e^{i\phi} & -\Delta),
\end{equation}
where $\Delta = f_{0} - f_{mw}$, $f_{0} = \gamma_{e} B_{0}$, $\Omega = \frac{\gamma_{e} B_{mw}}{2}$.
The eigenenergy $E_{\pm}$ of this effective Hamiltonian is
\begin{equation}
    E_{\pm} = \pm \frac{h}{2} \sqrt{\Delta^2 + \Omega^2}.
\end{equation}
The corresponding eigenstates are called dressed states. The resonance frequency of the system is therefore shifted by $\sqrt{\Delta^2 + \Omega^2} - \Delta$ by the application of microwave, and this effect is called AC Zeeman effect. \par
Note that, although an NV center can be regarded as an effective two-level system when a certain amount of static magnetic field is applied in the axial direction, it is actually three-level systems with spin $S = 1$. Therefore it is necessary to consider the difference caused by this gap. In this experiment, a modification is necessary with respect to $\Omega$. For a three-level system, $\Omega = \frac{\gamma_{e} B_{mw}}{\sqrt{2}}$ holds.

\subsection*{Rigorous time evolution by pulse sequence}
First, we derive the general time evolution of a two-level system under microwave irradiation \cite{ramsey1955resonance}. Under equation Eq. (\ref{eqa1}). When the initial state at the time $t = 0$ is
\begin{equation}
    \psi(0) = C_{0} \Ket{+} + C_{1} \Ket{-},
\end{equation}
The state at the time $t = T$ is
\begin{multline}
    \psi(T) = \Biggl\{ C_{0} \qty(\cos \pi \qty(\sqrt{\Delta^2 + \Omega^2} T) - i \cos \theta \sin \qty(\pi \sqrt{\Delta^2 + \Omega^2} T)) - i C_{1} \sin \theta e^{-i\phi} \sin \qty( \pi \sqrt{\Delta^2 + \Omega^2} T) \Biggr\} e^{- i\pi f_{\mathrm{mw}} T} \Ket{+} \\ 
     +  \Biggl\{ -iC_{0} \sin \theta e^{i\phi} \sin \qty(\pi \sqrt{\Delta^2 + \Omega^2} T) + C_{1} \qty(\cos \qty(\pi \sqrt{\Delta^2 + \Omega^2} T) + i \cos \theta \sin \qty(\sqrt{\Delta^2 + \Omega^2} T))  \Biggr\} e^{i \pi f_{\mathrm{mw}} T} \Ket{-},
\end{multline}
where $\cos \theta = \frac{\Delta}{\sqrt{\Delta^2 + \Omega^2}}$, $\sin \theta = \frac{\Omega}{\sqrt{\Delta^2 + \Omega^2}}$.
We use this formula and derive the time evolution of the pulse sequence. \par
\begin{figure}[H]
    \centering
    \includegraphics[scale=0.6]{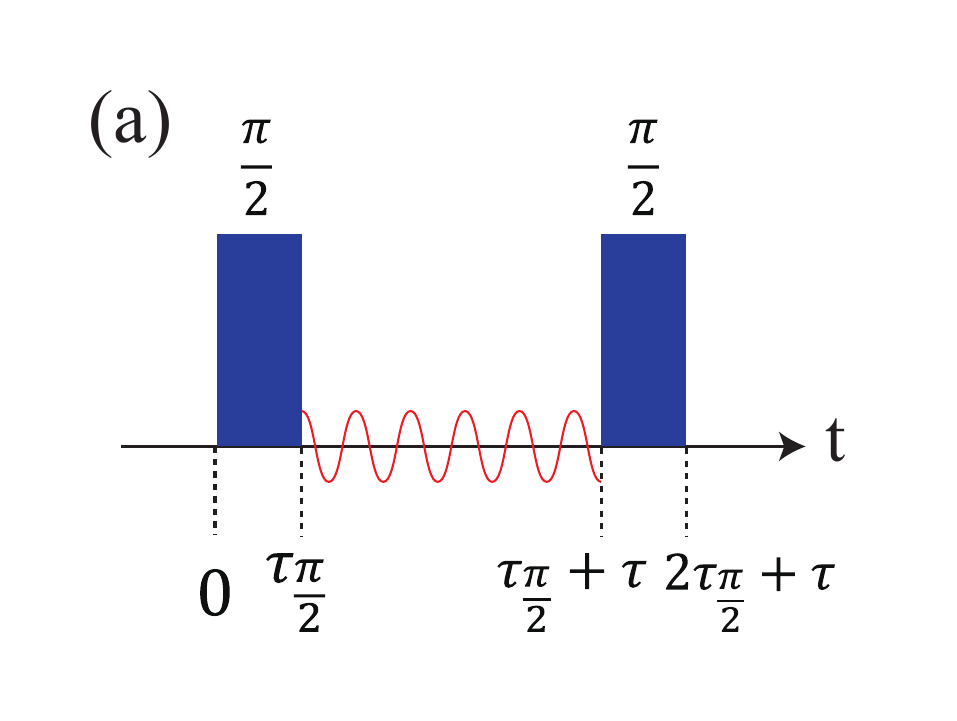}
    \caption{(a) Pulse protocol for detecting the AC Zeeman effect.}
    \label{fig:sb}
\end{figure}

 
The pulse sequence is shown in Fig. \ref{fig:sb}(a). Starting from the initial state $\psi(0) = \Ket{-}$, the following operations are performed. \\
\\
\noindent
(1) $\frac{\pi}{2}$ pulse operation with resonant microwave whose frequency is $f_{\text{mw}} = f_{0}$ (from $t = 0$ to $t = \tau_{\frac{\pi}{2}}$). \\
(2) off-resonant microwave ($f_{\text{mw}} = f_{s}$) irradiation (from $t = \tau_{\frac{\pi}{2}}$ to $t = \tau_{\frac{\pi}{2}} + \tau$). \\
(3) $\frac{\pi}{2}$ pulse operation with resonant microwave (from $t = \tau_{\frac{\pi}{2}} + \tau$ to $t = 2\tau_{\frac{\pi}{2}} + \tau$). \\

We let the initial phase of the resonant microwave be $0$ and define the phase of the off-resonant microwave at time $t = \tau_{\frac{\pi}{2}}$ as $\phi$. We calculate the state after each operation. \\
\\
\noindent
The state after operation (1) is
\begin{equation}
    \psi(\tau_{\frac{\pi}{2}}) = \frac{1}{\sqrt{2}} \qty(- ie^{-i \pi f_{0} \tau_{\frac{\pi}{2}}} \Ket{+} + e^{i\pi f_{0} \tau_{\frac{\pi}{2}}} \Ket{-}). 
\end{equation}
The state after operation (2) is
\begin{multline}
    \psi(\tau_{\frac{\pi}{2}} + \tau) = \frac{1}{\sqrt{2}} \Biggl\{ -\qty(i\cos \qty( \pi \sqrt{\Delta^2 + \Omega^2} \tau) + \cos \theta \sin \qty(\pi \sqrt{\Delta^2 + \Omega^2} \tau))e^{-i \pi f_{0} \tau_{\frac{\pi}{2}}} \\
     - i \sin \theta e^{-i\phi} \sin \qty( \pi \sqrt{\Delta^2 + \Omega^2} \tau) e^{i \pi f_{0} \tau_{\frac{\pi}{2}}} \Biggr\} e^{-i \pi f_{mw} \tau} \Ket{+} \\ 
     + \frac{1}{\sqrt{2}} \Biggl\{ - \sin \theta e^{i\phi} \sin \qty(\sqrt{\Delta^2 + \Omega^2} \tau) e^{-i \pi f_{0} \tau_{\frac{\pi}{2}}} + \qty(\cos \qty(\sqrt{\Delta^2 + \Omega^2} \tau) + i \cos \theta \sin \qty(\sqrt{\Delta^2 + \Omega^2} \tau)) e^{i \pi f_{0} \tau_{\frac{\pi}{2}}} \Biggr\} e^{i \pi f_{mw} \tau} \Ket{-}. 
\end{multline}
The state after operation (3) is
\begin{multline}
    \psi(2\tau_{\frac{\pi}{2}} + \tau) = -\Biggl\{ i \qty(\cos \frac{\Delta \tau}{2} \cos \qty(\sqrt{\Delta^2 + \Omega^2} \tau) + \cos \theta \sin \frac{\Delta \tau}{2} \sin \qty(\sqrt{\Delta^2 + \Omega^2} \tau)) \\
    + \sin \theta \sin (-\frac{\Delta \tau}{2} - \omega_{0} \tau_{\frac{\pi}{2}} + \phi) \sin \qty(\sqrt{\Delta^2 + \Omega^2} \tau) \Biggr\} e^{-2\pi i f_{0} \tau_{\frac{\pi}{2}}} e^{-i \pi f_{0} \tau} \Ket{+} \\ 
    + \Biggl\{ -i \qty(\sin \frac{\Delta \tau}{2} \cos \qty( \sqrt{\Delta^2 + \Omega^2} \tau) - \cos \theta \cos \frac{\Delta \tau}{2} \sin \qty(\sqrt{\Delta^2 + \Omega^2} \tau)) \\
    - \sin \theta \cos (-\frac{\Delta \tau}{2} - \omega_{0} \tau_{\frac{\pi}{2}} + \phi) \sin \qty(\sqrt{\Delta^2 + \Omega^2} \tau) \Biggr\} e^{2\pi i f_{0} \tau_{\frac{\pi}{2}}} e^{i \pi f_{0} \tau} \Ket{-}. 
\end{multline}
Therefore, the probability $P_{+}(2\tau_{\frac{\pi}{2}} + \tau)$ of a transition to the $\Ket{+}$ state by the pulse sequence can be written as
\begin{multline}
    P_{+}(2\tau_{\frac{\pi}{2}} + \tau) = \qty(\cos \frac{\Delta \tau}{2} \cos \qty(\sqrt{\Delta^2 + \Omega^2} \tau) + \cos \theta \sin \frac{\Delta \tau}{2} \sin \qty(\sqrt{\Delta^2 + \Omega^2} \tau))^{2}  \\
     \quad + \sin^2 \theta \sin^2 \qty(-\frac{\Delta \tau}{2} - \omega_{0} \tau_{\frac{\pi}{2}} + \phi) \sin ^2 \qty(\sqrt{\Delta^2 + \Omega^2} \tau)
\end{multline}
Let us consider the situation where the detuning $\Delta$ is sufficiently large compared to the signal microwave amplitude $\Omega$. In this case, we can approximate $\sin \theta \sim 0$, $\cos \theta \sim 1$ and the transition probability can be approximated as
\begin{align}
    P_{+}(2\tau_{\frac{\pi}{2}} + \tau) &= \frac{\cos \qty(2\pi (\sqrt{\Delta^2 + \Omega^2} - \Delta)\tau) + 1}{2} \\
    \label{eqa:prob}
    & \sim \frac{1 + \cos \qty(\pi \frac{\Omega^2}{\Delta} \tau)}{2}.
\end{align}
In our experiments, we invert the phase of the final $\frac{\pi}{2}$ pulse by $\pi$. Therefore, Eq. (\ref{eqa:prob}) corresponds to the probability of being in the $\Ket{-}$ state. Let $\alpha$ and $\beta$ be the counts in one frame of the CMOS camera of the $\Ket{-}$ and $\Ket{+}$ states respectively ($\alpha > \beta$), and let their ratio be $C = \frac{\alpha - \beta}{\alpha}$. \\
The signal $S(\tau)$ at microwave irradiation time $\tau$ can be written as
\begin{equation}
    S(\tau) = 1 + \frac{\cos \qty(\pi \frac{\Omega^2}{\Delta}\tau) - 1}{2} C.
\end{equation}
We add here a term due to relaxation. Considering that the total pulse sequence length is twice the microwave irradiation time, the signal can be written as 
\begin{equation}
    S(\tau) = 1 + \frac{\cos \qty(\pi \frac{\Omega^2}{\Delta}\tau) e^{\frac{-2\tau}{T_{2}}} - 1}{2} C.
\end{equation}

\subsection*{Time evolution with multiple $\pi$ pulses}
\begin{figure}[H]
    \centering
    \includegraphics[scale=0.6]{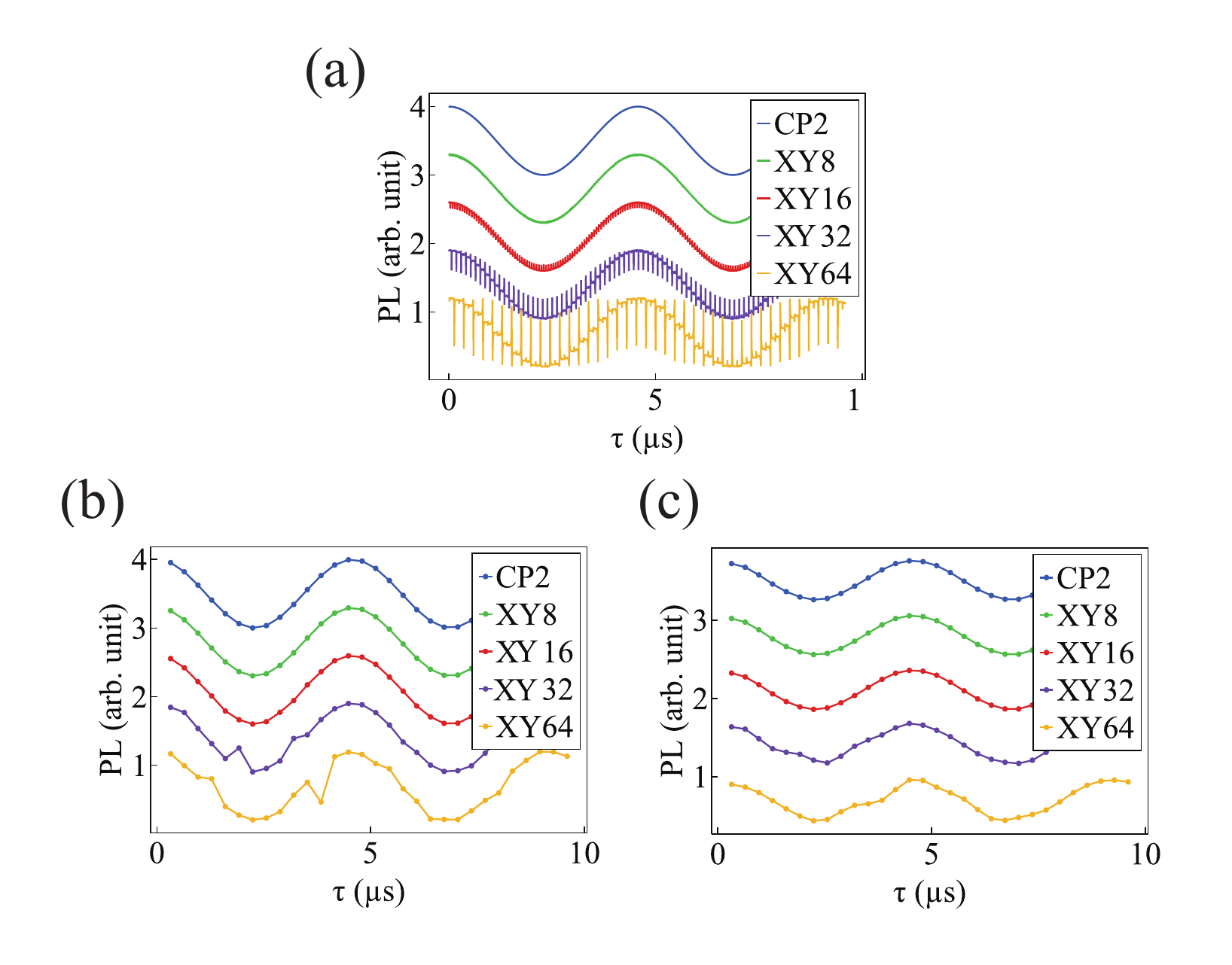}
    \caption{(a) Numerical calculations of time evolution with multiple $\pi$ pulses. (b) Numerical calculations for only $\tau$ sampled in the experiment in Fig. 3(b) in the main text. (c) Results of low-pass filtering for the data in Fig. \ref{fig:sc}(b).}
    \label{fig:sc}
\end{figure}

In this study, in addition to the protocol shown in Fig. \ref{fig:sb}(a), we develop a protocol in which multiple $\pi$ pulses are applied to extend the coherence time. We numerically calculate and discuss the time evolution of each sequence. Figure \ref{fig:sc}(a) shows the results corresponding to the number of $\pi$ pulses and parameters ($\Omega = 7.76 \, \mathrm{MHz}$, $\Delta = 140 \, \mathrm{MHz}$) in Fig. 3(b) in the main text. The calculations are performed for each operation using Eq. (S9). Because the initial phase of the signal microwave is random in actual experiments, we increment the phase of the off-resonant microwave from 0 to $2\pi$ at 0.01 intervals and average all of the results. Figure \ref{fig:sc}(a) tells that oscillations due to the AC Zeeman effect can be observed at all protocols. In addition, a comb-like modulation becomes more pronounced as the number of $\pi$ pulses increases. This modulation arises from the fact that the eigenstates of the effective Hamiltonian in a rotating frame, as given in Eq. (S6), consist of a slight mixture of the $m_{\text{s}} = 0$ and $m_{\text{s}} = -1$ state. Consequently, after the $\frac{\pi}{2}$ pulse operation, in addition to the phase accumulation due to the AC Zeeman effect, there exists a state transition process. The frequency of this state transition is $\sqrt{\Delta^2 + \Omega^2}$, which is the energy difference between the dressed eigenstates divided by Planck's constant, and the modulation is maximized when the interval between $\pi$ pulses is half-integer times this period. Since this modulation affects only a limited periodic range of $\tau$, the effect is small in actual measurements with rough sweeps of $\tau$. Figure \ref{fig:sc}(b) shows the numerical calculations for only $\tau$ sampled in our experiment. Although some modulation exists, no large comb-like modulation is observed. Also, since this modulation is of high frequency and the signal due to the AC Zeeman effect is of low frequency, it is possible to reduce the effect by cutting the high frequency component of the signal. Figure \ref{fig:sc}(c) shows the low frequency component extracted from the simulation in Fig. \ref{fig:sc}(b) with the cutoff frequency of $1 \, \mathrm{MHz}$, which confirms that the modulation is significantly suppressed, especially in the $\mathrm{XY32}$ and $\mathrm{XY64}$ protocols. \par
Since no significant high-frequency component is observed in the experimental data in our paper even when the number of $\pi$ pulse is increased, we analyze our data without low-pass processing.

\section*{C. Composite pulse sequences}
When conducting imaging measurements, it is crucial to irradiate microwave for pulse controls both strongly and uniformly within the field of view. However, there are many cases where achieving this requirement is challenging, and our experiment is no exception. For example, while the antenna depicted in Fig. 1(b) in the main text can emit a strong and uniform microwave within the field of view, it is known that when a metal pattern is present, it couples to the microwave antenna inductively and intensifies microwave amplitude in its vicinity \cite{Mariani2020}. Thus, in this study, we adopt a strategy that uses a sequence robust against pulse length errors, with some distribution of microwave amplitude within the field of view. As methods for constructing pulse sequences that are robust against pulse errors, techniques such as composite pulses or GRAPE and other methods from optimal control have been employed \cite{jones2011quantum} and have been applied to wide-field measurements using NV centers \cite{Ziem2019,nomura2021composite}. In our experiment, we use composite pulses, particularly SCROFULOUS composite pulse sequence \cite{cummins2003tackling} because it has an advantage of having a shorter total pulse length.

\section*{D. Sensitivity Estimation}
After irradiating non-resonant microwaves for time $\tau_{i} \ (i = 1, 2, ... , k)$, the signal $S_{i}$ can be written as
\begin{equation}
     S_{i} =  S(\tau_{i}) = 1 + \frac{\cos \qty(\pi \frac{\Omega^2}{\Delta}\tau_{i}) e^{\frac{-2\tau_{i}}{T_{2}}} - 1}{2} C,
\end{equation}
and we obtain the data with noise added to this signal in our measurements. From this equation, we derive the sensitivity of the microwave field $B_{\mathrm{mw}}$ of the signal microwave. \par
We let $\sigma^{2}(T)$ be the variance of each signal due to noise at integration time $T$. This variance should be mainly due to optical shot noise and is considered to scale with $T^{-\frac{1}{2}}$. The covariance matrix of the parameter $\hat{\theta}$ in nonlinear regression is generally expressed as
\begin{equation}
    \mathrm{COV}(\hat{\theta}) = \sigma^{2} \qty(J^{T}J)^{-1},
    \label{eqc:sig}
\end{equation}
where $J$ is Jacobian, whose component $J_{ij}$ can be written as
\begin{equation}
    J_{ij} = \frac{\partial S_{i}}{\partial \hat{\theta_{j}}}.
\end{equation}
In this experiment, we evaluate only $B_{\mathrm{mw}}$, so $\hat{\theta}$ is taken one-dimensional. Based on Eq. (\ref{eqc:sig}), the variance of $B_{\mathrm{mw}}$ can be evaluated as 
\begin{equation}
    \sigma^{2}_{B_{\mathrm{mw}}} = \frac{\sigma^{2}}{\sum_{i = 1}^{k} J_{i}^{2}},
\end{equation}
\begin{equation}
    J_{i} = \frac{\partial S_{i}}{\partial B_{\text{mw}}} = \frac{\gamma_{e} \pi \frac{\Omega}{\Delta} \tau_{i} \sin \qty(\pi \frac{\Omega^2}{\Delta}\tau_{i}) e^{\frac{-2\tau_{i}}{T_{2}}}}{\sqrt{2}}C.
\end{equation}
We evaluate $\sigma^2$ using the residuals from the fitting results to the signal integrated for about one hour.

\section*{E. Relationship between the number of $\pi$ pulses and $T_{2}$}
\begin{figure}[H]
    \centering
    \includegraphics[width=\linewidth]{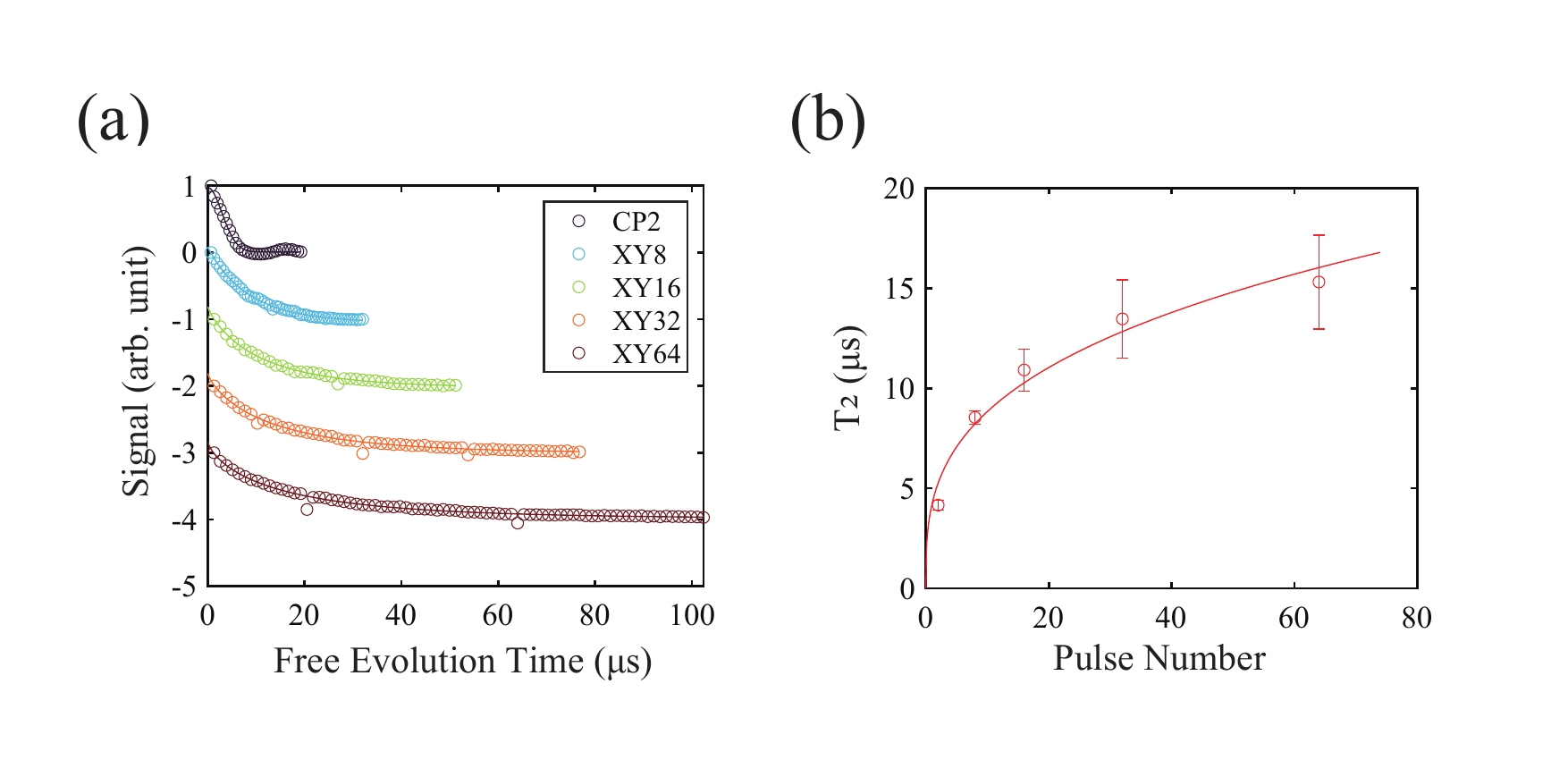}
    \caption{(a) Signals for each pulse sequence. (b) Relationship between $\pi$ pulse number and $T_{2}.$}
    \label{fig:sd}
\end{figure}

We evaluate the transverse relaxation time $T_{2}$ by applying $\mathrm{CP2}$ sequence or $\mathrm{XY8}$ sequences and investigate the relationship between $\pi$ pulse number and $T_{2}$ of our sample. Figure \ref{fig:sd}(a) shows the signals from each sequence. The periodic dips observed in the $\mathrm{XY8}$ sequences are due to the effect of $^{15}\mathrm{N}$ caused by a slight misalignment of the external magnetic field. The signals are fitted by $\exp(-\frac{\tau}{T_{2}})$, and the obtained $\pi$ pulse number dependence of $T_{2}$ is shown in Fig. \ref{fig:sd}(b). It can be observed that $T_{2}$ extends as the $\pi$ pulse number $N_{\pi}$ increases. The scaling factor of $T_{2} \propto N_{\pi}^s$ is determined to be $s = 0.41$ by fitting.

\section*{F. Scaling factor of the sensitivity's dependence on the number of $\pi$ pulses}
We discuss the validity of the scaling factor $p$ of the sensitivity's dependence on the number of $\pi$ pulses. We obtain the scaling factor as $p = 0.98$ by fitting the result in Fig. 3(d) with Eq. (4) in the main text. Theoretically, sensitivity primarily depends on $T_{2}$ and the microwave irradiation time $\tau$. The best sensitivity $\eta_{\mathrm{best}}$ is given in Eqs. (4) and (5) and given that the readout time is sufficiently longer than coherence time, $\eta_{\mathrm{best}} \propto T_{2}^{-1}$ is expected. The NV ensembles used in our experiment is measured to have a dependence of $T_{2} \propto N_{\pi}^{0.41}$ Therefore, if the measurement is conducted at a single microwave irradiation time with the best sensitivity, it is expected to have a dependence of $\eta \propto N_{\pi}^{-0.41}$ with $p = 0.41$ instead of $0.98$. We attribute this difference to the sampling conditions. In practice, we sweep $\tau$ to observe oscillations. When the $\pi$ pulse number is small and the coherence time is short, the sensitivity is much worse than the best because the signal is fully attenuated at many points and has no sensitivity. Therefore, the experimental scaling factor $p$ is expected to be larger than the ideal one estimated only from the coherence time. \par

\section*{G. Strategies for optimizing the experimental setup to enhance sensitivity}
The sensitivity of our sequence can be further enhanced by optimizing the experimental setup. First, we can shorten the long readout time $\tau_{\text{read}} \, (= 64 \, \mathrm{\mu s})$ which limits the sensitivity by increasing the optical laser power. In this study, the optical laser power is set to 150 mW, and by tripling it, for example, it is possible to reduce $\tau_{\text{read}}$ to one-third while maintaining the PL intensity. Second, we can devise the sample preparation. In our study, we have created NV centers at a depth of about 30 nm by 10 keV ion implantation. However, in imaging measurements, the spatial resolution is limited by the optical resolution, and it does not make a significant difference if we create NV ensembles at a deeper position. By simultaneously increasing the implantation energy and dose, the number of NV centers per pixel can be increased while maintaining the defect density. For example, the condition, 100 keV with $4 \times 10^{13} \, \text{cm}^{-2}$ \cite{healey2020comparison}, could quadruple the PL intensity without degrading both the spatial resolution and the coherence time. In addition, the thickness of the diamond sample used in this study is $500 \, \mathrm{\mu m}$, and PL is read out from the backside. Due to the high refractive index of diamond, optical aberrations result in about one order of magnitude reduction in PL intensity compared to a thinner diamond sample. Therefore, by reducing the thickness of the diamond sample (e.g., $30 \, \mathrm{\mu m}$ thickness), the PL intensity is expected to increase by an order of magnitude. With these strategies, the sensitivity can be increased by at least an order of magnitude.

\section*{H. Comparison of the sensitivity with previous research}
We compare the sensitivity obtained in our experiment with the values from previous research. There are several papers that have explored microwave sensing using NV centers and evaluated sensitivity. For instance, Ref. \cite{Alsid2023} reports a record sensitivity of $3.4 \, \text{pT}/\sqrt{\text{Hz}}$ for resonant microwaves using an ensemble of NV centers. 
However, the sensitivity in many papers, including Ref. \cite{Alsid2023} is that of resonant microwave based on Rabi oscillation measurement. Since the protocol is significantly different from our study, it is not fair to directly compare the obtained sensitivity. \par
Thus, we compare the sensitivity with Ref. \cite{li2019wideband}, which measured off-resonant microwave using a protocol similar to our study. The sensitivity of off-resonant microwave amplitude largely depends on the detuning and microwave amplitude, making it impossible to directly compare the sensitivity obtained in experiments. However, it is possible to calculate the best sensitivity by substituting the experimental parameters of our experiment and microwave detuning $\Delta = 4 \, \text{MHz}$ and amplitude $B_{\text{mw}} = 25 \, \mathrm{\mu T}$ used for evaluation of sensitivity in Ref. \cite{li2019wideband} into Eqs. (5) and (6) in the main text. As a result, we obtain a sensitivity of $11.8 \, \mathrm{\mu T} / \sqrt{\text{Hz}}$ for the protocol using the CP2 sequence. This value is worse than the sensitivity of $0.43 \, \mathrm{\mu T} / \sqrt{\text{Hz}}$ in Ref. \cite{li2019wideband}. In Ref. \cite{li2019wideband}, the experiment is conducted using a confocal microscope system with a single NV center that has a quite long coherence time $T_{2}$, whereas we use a wide-field microscope system with an ensemble of NV centers that has a shorter $T_{2}$. The differences in the coherence time and quantum efficiency of the optical system are the main reasons for the difference in the sensitivity.